 \documentclass[11pt,twocolumn]{article}

 \pdfoutput=1

 \usepackage[english]{babel}
\usepackage{graphicx} 
\usepackage{amsmath}

\usepackage{enumitem}

\usepackage{float}
\usepackage{ifthen}
\usepackage{hyperref}

\usepackage{titlesec}

\titlespacing*{\section}{0pt}{0.2\baselineskip}{\baselineskip}

\begin{document}


\pagestyle{empty}


\twocolumn[
\begin{center}
{\bf \huge 
An improved approximation for the moist-air entropy potential temperature $\theta_s$.
}\\
\vspace*{3mm}
{\Large \bf by Pascal Marquet } {\Large (WGNE Blue-Book 2015)}. \\
\vspace*{2mm}
{M\'et\'eo-France. CNRM/GMAP.
 Toulouse. France.}
{\it E-mail: pascal.marquet@meteo.fr} \\
\vspace*{1mm}
\end{center}
]





 \section{\underline{\Large Motivations}} 
\vspace{-4mm}

The moist-air entropy is defined in Marquet (2011, hereafter M11) by $\boxed{ s  =  s_{ref}  + c_{pd} \: \ln(\theta_{s}) }$ in terms of two constant values ($s_{ref}$, $c_{pd}$) and a potential entropy temperature denoted by $\theta_s$.
It is shown in M11 that a quantity denoted by $({\theta}_{s})_1$ plays the role of a leading order approximation of ${\theta}_{s}$.

The aim of this note is to demonstrate in a more rigorous way that $ ({\theta}_{s})_1$ is indeed the leading order approximation of ${\theta}_{s}$, and to derive a second order approximation which may be used in computations of values, gradients or turbulent fluxes of moist-air entropy.
Some impacts of this second order approximation are described in this brief version of a note to be submitted to the QJRMS.

 \section{\underline{\Large Definition of $\theta_s$ and $(\theta_s)_1$}} 
\label{section2}
\vspace{-4mm}

The potential temperature $\theta_s$ is defined in M11 by
\vspace*{-2mm}
\begin{equation}
\! \! \boxed{
  {\theta}_{s}    =  ({\theta}_{s})_1  
        \left( \frac{T}{T_r}\right)^{\!\!\lambda \,q_t}
 \!  \! \left( \frac{p}{p_r}\right)^{\!\!-\kappa \,\delta \,q_t}
 \!  \! \left( \frac{r_r}{r_v} \right)^{\!\!\gamma\,q_t}
      \frac{(1\!+\!\eta\,r_v)^{\,\kappa \, (1+\,\delta \,q_t)}}
           {(1\!+\!\eta\,r_r)^{\,\kappa \,\delta \,q_t}}
  }
 \nonumber
\end{equation}
\vspace*{-3mm} 
\begin{equation} 
\mbox{where} \; \; 
  \boxed{
    ({\theta}_{s})_1 =   \theta_l \;  \exp\! \left(  \Lambda_r \: q_t  \right)
  \:  .}
\end{equation}

\vspace*{-2mm} 
\noindent This definition of ${\theta}_{s}$ is rather complex, but the more simple quantity $({\theta}_{s})_1$ was considered in M11 as a leading order approximation of ${\theta}_{s}$, where 
the Betts' potential temperature is written as $ {\theta}_l  =  \theta \; 
 \exp [ \:  - \:   ({L_v\:q_l + L_s\:q_i}) \, / \, ({{c}_{pd}\:T})  \: ]$.

The total water specific content is $q_t = q_v + q_l + q_i$ and $r_v$ is the water vapour mixing ratio.
Other thermodynamic constants are: 
$R_d \approx 287$~J~K${}^{-1}$, 
$R_v \approx 461.5$~J~K${}^{-1}$, 
${c}_{pd}\approx 1005$~J~K${}^{-1}$, 
${c}_{pv}\approx 1846$~J~K${}^{-1}$, 
$\kappa = R_d/{c}_{pd} \approx 0.286$, 
$\lambda = {c}_{pv}/{c}_{pd} - 1 \approx 0.838$, 
$\delta = R_v/R_d - 1 \approx 0.608$, 
$\eta = R_v/R_d \approx 1.608$, , 
$\varepsilon = R_d/R_v \approx 0.622$ and 
$\gamma = R_v/{c}_{pd} \approx 0.46$.

The  term
\vspace*{-3mm} 
\begin{equation} 
\boxed{ \;
\Lambda_r = \left[  \: (s_{v})_r - (s_{d})_r \: \right] \, / \, c_{pd} \; \approx \; 5.87
\; }
\label{def_Lambda_r}
\end{equation}

\vspace*{-4mm}
\noindent depends on reference entropies 
of dry air and water vapour at $T_r=0$~C, denoted by 
$(s_{d})_r = s_{d}(T_r, e_r)$ and $(s_{v})_r = s_{v}(T_r, p_r-e_r)$, where
$e_r = 6.11$~hPa is the saturating pressure at $T_r$ and $p_r = 1000$~hPa. 
The two reference entropies 
$(s_{v})_r \approx 12673 $~J~K${}^{-1}$ and 
$(s_{d})_r \approx  6777$~J~K${}^{-1}$
are computed in M11 from the Third Law of thermodynamics, leading to
$\Lambda_r \approx 5.87$.
The reference mixing ratio is $r_r = \varepsilon \: e_r \, / \,  (p_r - e_r) \approx 3.82 $~g~kg${}^{-1}$.

 \section{\underline{\Large Computations of  $\Lambda_{s}$ and $\Lambda_{\star}$}} 
\vspace{-4mm}

Let us define $\Lambda_{s}$ by the formula ${\theta}_{s}=   \theta_l \;  \exp\! \left(  \Lambda_s \: q_t  \right)$ where  ${\theta}_{s}$, ${\theta}_{l}$ and $q_t$ are known quantities, and thus by

\vspace*{-4.5mm} 
\begin{equation} 
\Lambda_s  = \frac{1}{q_t} \; \ln\!\left(  \frac{{\theta}_{s}}{{\theta}_{l}}   \right).
\label{def_lambda_s}
\end{equation}

\vspace*{-2.5mm}
\noindent The aim is to compute $\Lambda_s$ from (\ref{def_lambda_s}) for a series of 16 vertical profiles of stratocumulus and cumulus with varying values of ${q_t}$, ${\theta}_{s}$ and ${\theta}_{l}$, in order to analyse the discrepancy of $\Lambda_s$ from the constant value $\Lambda_r \approx 5.87$  given by (\ref{def_Lambda_r}).

\begin{figure}[hbt]
\centering
\includegraphics[width=0.99\linewidth]{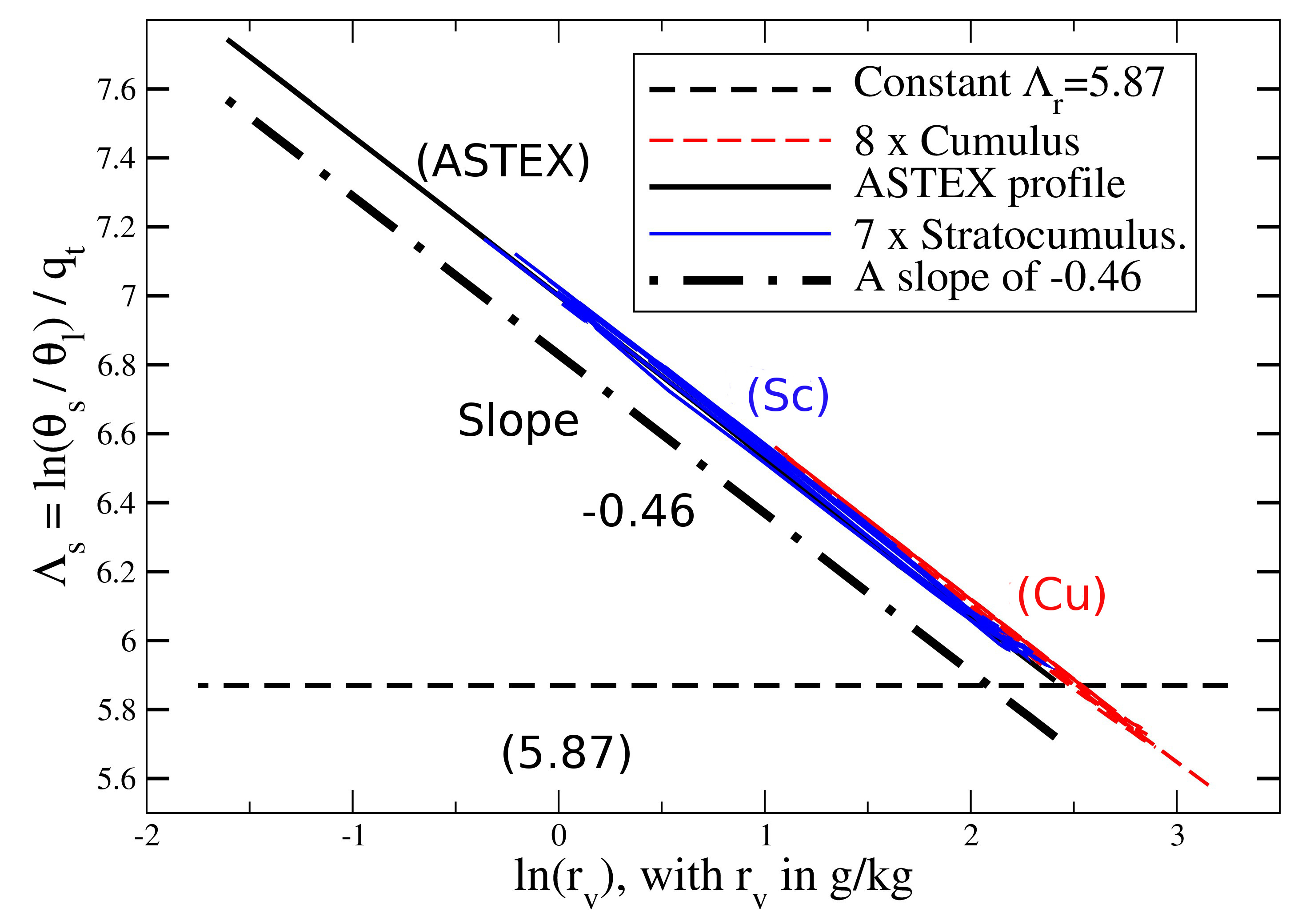}
\vspace{-10mm}
\caption{\small \it 
A plot of $\Lambda_s$ against $\ln(r_v)$ for 8 cumulus (dashed red), 7 stratocumulus (solid blue) and ASTEX (solid black) vertical profiles.
The constant value $\Lambda_r \approx 5.87$ corresponds to the horizontal dashed black line.
An arbitrary line with a slope of $-0.46$ is added in dashed-dotted thick black line
\label{fig_1}}
\end{figure}

A trial and error process has shown that plotting $\Lambda_s$ against $\ln(r_v)$ leads to relevant results (see Fig.\ref{fig_1}).
Clearly, all stratocumulus and cumulus profiles are nearly aligned along a straight line with a slope of about $- 0.46$.
It is likely that this slope must correspond to $- \, \gamma$.
This linear law appears to be valid for a large range of $r_v$ (from $0.2$ to $24$~g~kg${}^{-1}$).
\vspace*{-4mm} 

It is then useful to find a mixing ratio $r_{\star}$ for which
\vspace*{-1.5mm} 
\begin{align} 
\Lambda_s  
  \; \approx \; 
\Lambda_{\star} 
& \;  = \;
5.87 \: - \: 0.46 \: \ln\,( r_v / r_{\star})
\label{def_lambda_star1} \:
\end{align}

\begin{figure}[hbt]
\centering
\includegraphics[width=0.99\linewidth]{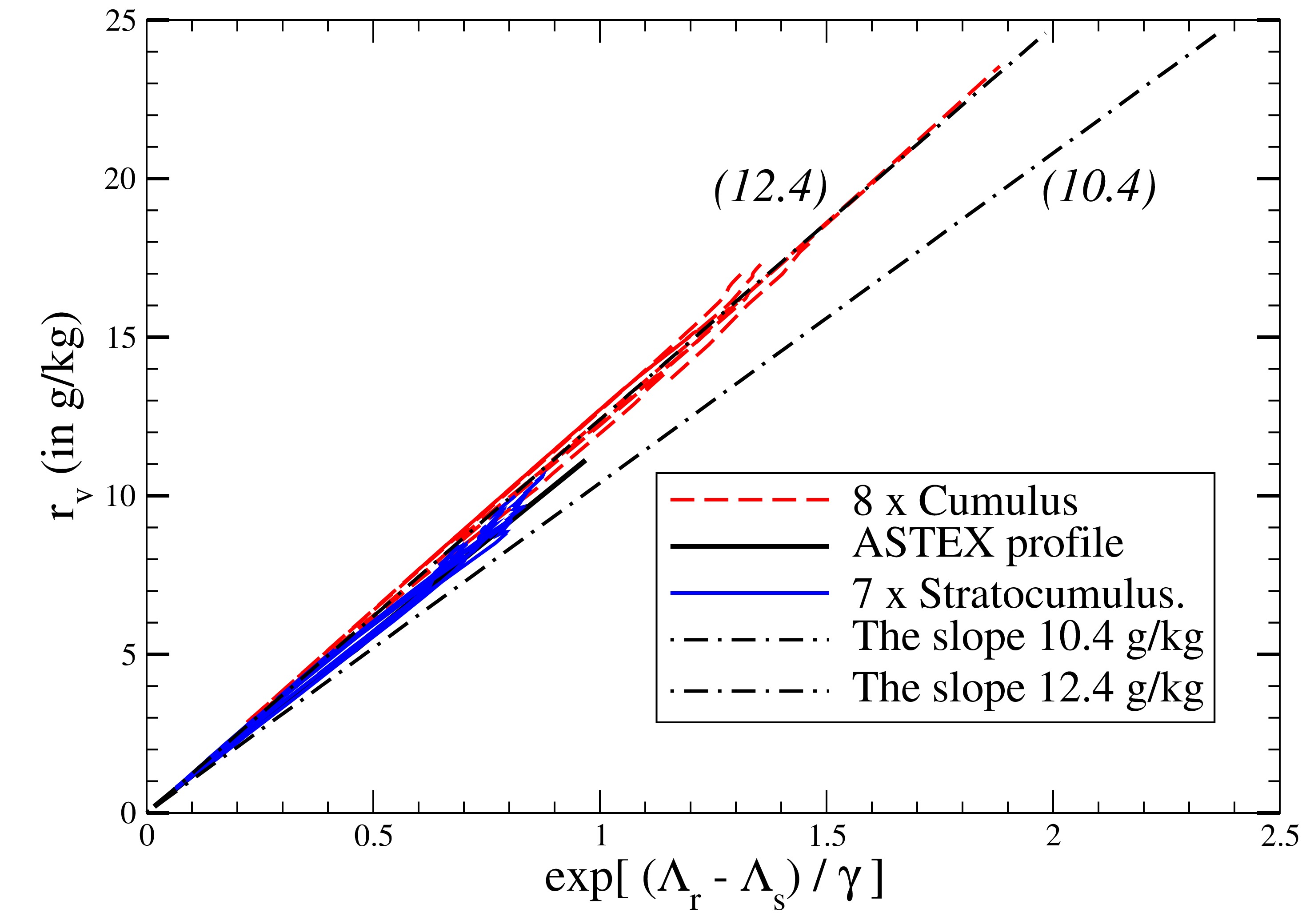}
\vspace{-9mm}
\caption{\small \it 
Same as Fig.\ref{fig_1}, but with $r_v$ plotted against the quantity $\exp[\: (\Lambda_r - \Lambda_s)/\gamma \:]$.
Two lines of slope $r_{\ast} = 10.4$ and $12.4$~g~kg${}^{-1}$ are added as dashed-dotted thin black lines.
\label{fig_2}}
\end{figure}

\vspace*{-3mm}
\noindent hold true, where $r_{\star}$ will play the role of positioning the dashed-dotted thick black line of slope $- \,\gamma \approx - 0.46$ in between the cumulus and stratocumulus profiles.
This corresponds to a linear fitting of  $r_v$ against $\exp[\: (\Lambda_r - \Lambda_s)/\gamma \:]$, $r_{\star}$ being the average slope of the scattered data points.
It is shown in Fig.\ref{fig_2} that the value $r_{\star} \approx 12.4$~g~kg${}^{-1}$ corresponds to a relevant fitting of all cumulus and stratocumulus vertical profiles for a range of $r_v$ up to $24$~g~kg${}^{-1}$.
\vspace*{-14mm} 

\begin{figure}[hbt]
\centering
\includegraphics[width=0.99\linewidth]{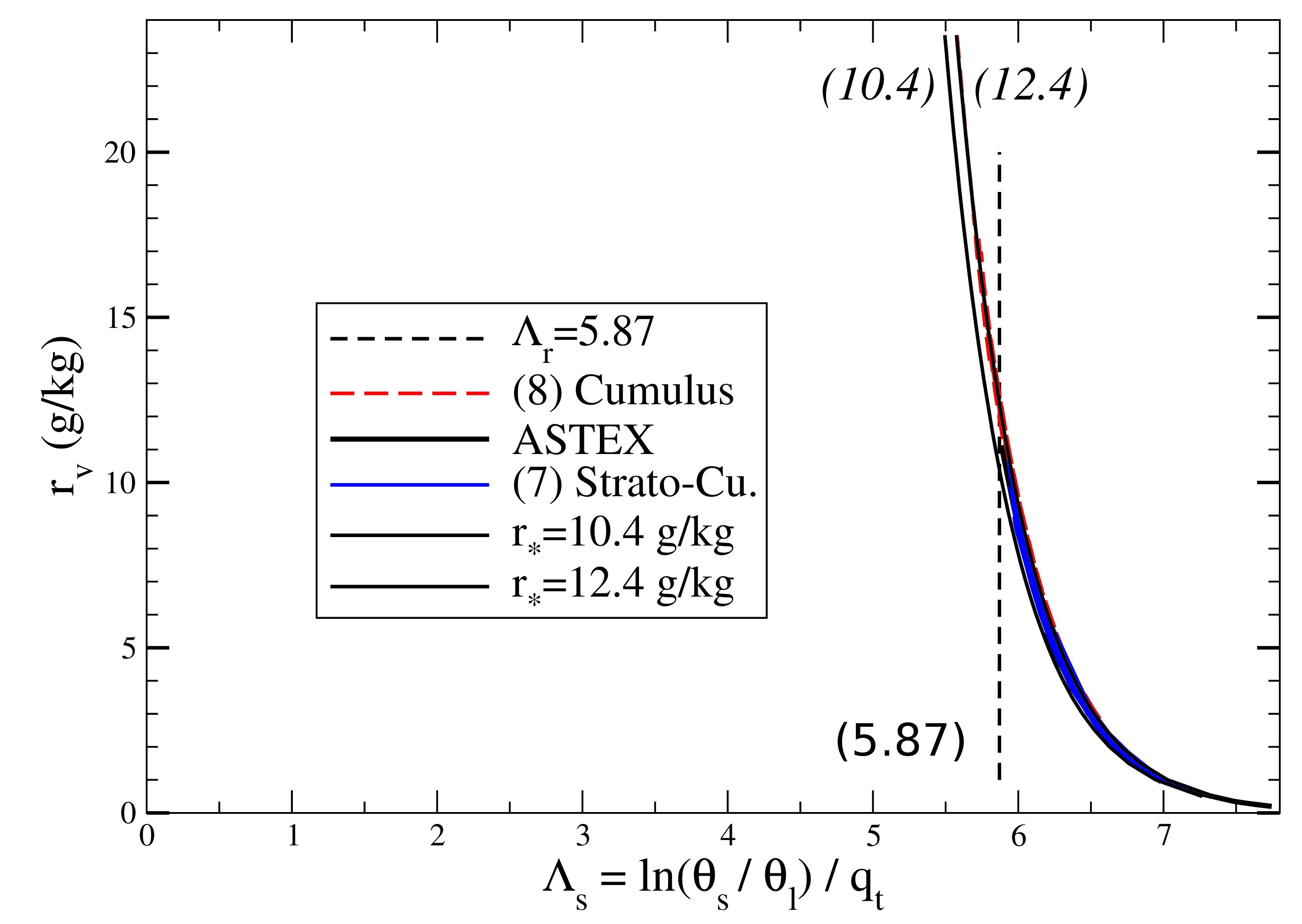}
\vspace{-9mm}
\caption{\small \it 
Same as Fig.\ref{fig_1}, but with $r_v$ plotted against the quantity $\Lambda_s$.
The two thin black lines correspond to (\ref{def_lambda_star1}) with $r_{\ast} = 10.4$ or $12.4$~g~kg${}^{-1}$.
\label{fig_3}}
\end{figure}

It is shown in Fig.\ref{fig_3} that  $\Lambda_s$ can indeed be approximated by $\Lambda_{\star}(r_v, r_{\ast})$ given by (\ref{def_lambda_star1}), with a clear improved accuracy in comparison with the constant value $\Lambda_r \approx 5.87$ for a range of $r_v$ between $0.2$ and $24$~g~kg${}^{-1}$.
Curves of $\Lambda_{\star}(r_v, r_{\ast})$ with $r_{\ast} = 10.4$ and $12.4$~g~kg${}^{-1}$ (solid black lines) both simulate with a good accuracy the non-linear variation of $\Lambda_s$ with $r_v$, and both simulate the rapid increase of $\Lambda_s$ for $r_v < 2$~g~kg${}^{-1}$.
\vspace{-5mm}

 \section{\underline{\Large Mathematical computation of $\Lambda_{\star}$}} 
\vspace{-4mm}

It is important to confirm, by using mathematical arguments, that $ ({\theta}_{s})_1$ corresponds to the leading order approximation of ${\theta}_{s}$, and that the slope of $- \,\gamma \approx - 0.46$ (with $r_{\ast} \approx 10.4$ or $12.4$~g~kg${}^{-1}$) corresponds to a relevant second order approximation for $\theta_s$.
These results are briefly mentioned in Marquet and Geleyn (2015).

First- and second-order approximations of $\theta_s$ can be derived by computing Taylor expansions for most of terms in the first formula recalled in Section~\ref{section2}.

The main result is that the term $(r_r/r_v)^{(\gamma\:q_t)}$ is exactly equal to $\exp[\: - \: ( \gamma \: q_t )\: \ln(r_v/r_r) \: ]$.

Then, the  first order expansion of  $(1 + \eta \: r_v)^{[\: \kappa \: (1+\delta \:q_t)\:]}$ for small $r_v \approx q_t$ is equal to $\exp(\: \gamma \: q_t )$, since $\gamma = \kappa \: \eta$.
The last term $(1 + \eta \: r_r)^{(\kappa \: \delta \:q_t)}$ leads to the higher order term $\exp[\: \gamma \:  \delta \: q_t \: r_r \: ] \approx \exp[\: O(q_t^2) \:]$, since $r_r  \ll 1$ and $q_t \ll 1$.
Other terms depending on temperature and pressure are exactly equal to $\exp[\: \lambda \; q_t \: \ln(T/T_r) \: ]$ and $\exp[\: - \: \kappa \: \delta \: q_t \: \ln(p/p_r) \: ]$.

The Taylor expansion of $\theta_s$ can thus be written as
\vspace{-2mm}
\begin{align}
\!\!\!\!\!\!\!\!\!\!\!\!
   & 
   \boxed{ \; {\theta}_{s}
     \approx  
         \: \theta \:
        \exp \left( - \:
                     \frac{L_{\mathrm{vap}}\:q_l + L_{\mathrm{sub}}\:q_i}{{c}_{pd}\:T}
                \right) 
      \; \exp \left( {\Lambda}_{\ast}  \: q_t \right)
      \; }
   \label{def_THs_approx} \\
   &  \; \; \; \;
    \boxed{ \times
      \; \exp \left\{ q_t  \left[ 
                \lambda \: \ln\!\left(\frac{T}{T_r} \right) 
                  - 
                 \kappa \: \delta \: \ln\!\left( \frac{p}{p_r} \right) 
                  \right]  
         + O(q_t^2) \:\right\}
      }
  \: , \nonumber
\end{align}

\vspace*{-6mm}
\begin{align}
\mbox{where} \; \; \; \;
\boxed{ \; {\Lambda}_{\ast} = {\Lambda}_r \: - \: \gamma \; \ln(r_v/r_{\ast}) \; }
 \: \nonumber
\end{align} 

\vspace*{-3mm}
\noindent 
and $r_{\ast} = r_r \times \exp(1) \approx 10.4$~g~kg${}^{-1}$ (see Figs.\ref{fig_2} and \ref{fig_3}).

The first order approximation of ${\theta}_{s}$ is thus given by the first line of (\ref{def_THs_approx}) with ${\Lambda}_{\ast} = {\Lambda}_r$, namely by the expected $({\theta}_{s})_1$.
An improved second order approximation is obtained by using ${\Lambda}_{\ast}$ instead of ${\Lambda}_r$ and  by taking into account the small corrective term $- \: \gamma \; \ln(r_v/r_{\ast})$.

Impacts of the second line of (\ref{def_THs_approx}) with terms depending on temperature and pressure lead to higher order terms which must explain the  fitted value $r_{\ast} \approx 12.4$~g~kg${}^{-1}$ observed for usual atmospheric conditions.

\vspace{-2mm}

 \section{\underline{\Large Conclusions}} 
\vspace{-5mm}

It has been shown that it is possible to justify mathematically the first order approximation of  $\theta_s$ derived in M11 and denoted by $(\theta_s)_1  \approx \theta_l \; \exp \left( {\Lambda}_r  \: q_t \right)$, which depends on the two Betts' variables $(\theta_l, q_t)$ and $\Lambda_r \approx 5.87$, leading to 
${ s  \approx  s_{ref}  + c_{pd} \: \ln(\theta_{l}) +c_{pd} \: \Lambda_r \: q_t }$ .

A second order approximation is derived and compared to observed vertical profiles of cumulus and stratocumulus, leading to
${ s  \approx  s_{ref}  + c_{pd} \: \ln(\theta_{l}) +c_{pd} \: \Lambda_{\ast} \: q_t }$, where the second order term ${ {\Lambda}_{\ast} = {\Lambda}_r \: - \: \gamma \; \ln(r_v/r_{\ast}) }$ depends on the constant $\Lambda_r \approx 5.87$, on the mixing ratio $r_v$, and on a tuning parameter $r_{\ast} \approx 12.4$~g~kg${}^{-1}$.

The use of the second order term ${\Lambda}_{\ast}$ depending on the non-conservative variable $r_v \approx q_t - q_l - q_i$ can explain why  it is needed to replace the Betts' potential temperature $\theta_l$ for computing flux of moist-air entropy: \\

\vspace*{-5mm}
\noindent $\overline{w'\theta'_s} \approx 
\exp({\Lambda}_{\ast}\:q_t) \: \overline{w' \theta'_l} 
+ {\Lambda}_{\ast}\:\theta_s \: \overline{w' q'_t}  
- (\gamma \, q_t \, \theta_s / r_v) \: \overline{w 'r'_v}$, \\
the last term depending on $\overline{w 'q'_t}$ minus $\overline{w 'q'_l}$ or $\overline{w 'q'_i}$.

\vspace{1mm}
\noindent{\large\bf \underline {References}}
\vspace{0mm}

\noindent{$\bullet$ Marquet P.} {(2011)}.
Definition of a moist entropic potential temperature. 
Application to FIRE-I data flights.
{\it Q. J. R. Meteorol. Soc.}
{\bf 137} (656) :
p.768--791.

\noindent{$\bullet$ Marquet P. and Geleyn, J.-F.} {(2015)}.
Formulations of moist thermodynamics for atmospheric modelling.
To appear in 
{\it Parameterization of Atmospheric Convection}, Volume~II
(R. S. Plant and J. I. Yano, Eds.), 
Imperial College Press, in press.


\end{document}